\documentclass[showpacs,preprintnumbers,preprint,aps,amsmath,amssymb]{revtex4}
\usepackage{amsmath, amssymb, graphics,epsfig}
\newcommand{\mathsym}[1]{{}}

\def\lsim{\:\raisebox{-1.1ex}{$\stackrel{\textstyle<}{\sim}$}\:}
\def\gsim{\:\raisebox{-1.1ex}{$\stackrel{\textstyle>}{\sim}$}\:}
\def\10{$SO(10)$}
\def\21{SU(2) $\otimes$ U(1) }

\def\422{$SU(4) \otimes SU(2) \otimes SU(2)$}
\def\321{SU(3) $\otimes$ SU(2) $\otimes$ U(1)}
\def\lsim{\raise0.3ex\hbox{$\;<$\kern-0.75em\raise-1.1ex\hbox{$\sim\;$}}}
\def\gsim{\raise0.3ex\hbox{$\;>$\kern-0.75em\raise-1.1ex\hbox{$\sim\;$}}}
\def\vev#1{\left\langle #1\right\rangle}

\newcommand{\ba}{\begin{array}}
\newcommand{\ea}{\end{array}}
\newcommand{\be}{\begin{equation}}
\newcommand{\ee}{\end{equation}}
\newcommand{\beqa}{\begin{eqnarray}}
\newcommand{\eeqa}{\end{eqnarray}}
\relax
\def\321{$SU(3)\times SU(2)\times U(1)$}
\def\mt{$\mu$-$\tau$ }

\begin{document}
\bigskip
\bigskip
\title{Type I seesaw mechanism for quasi degenerate neutrinos} 
\author{ Anjan S.  Joshipura\footnote{anjan@prl.res.in} , Ketan M. Patel  
\footnote{kmpatel@prl.res.in} } \affiliation{ Physical Research
Laboratory, Navarangpura, Ahmedabad 380 009, India }
\author{ Sudhir K. Vempati \footnote{vempati@cts.iisc.ernet.in} 
 } \affiliation{ Centre for High Energy Physics, Indian Inst. of Science, Bangalore 560 012, India \vskip 1.0truecm}

\begin{abstract} 
\vskip 1.0 truecm 
We discuss symmetries and scenarios leading to quasi-degenerate neutrinos in type-I seesaw models.  The existence of
degeneracy in the present approach is not linked to any specific structure for the Dirac neutrino Yukawa coupling matrix  $y_D$ and holds in general.  Basic input is the application of the minimal flavour violation principle to the leptonic sector. Generalizing this principle, we assume  that the structure of the right handed neutrino mass matrix is determined by $y_D$ and the charged lepton Yukawa coupling matrix $y_l$ in an effective theory  invariant under specific  groups ${\cal G}_F$ contained in the full symmetry group of the
kinetic energy terms. ${\cal G}_F$ invariance also leads to specific structure for the departure from degeneracy. The neutrino mass matrix (with degenerate mass $m_0$) resulting after seesaw mechanism has a simple form ${\cal M}_\nu\approx m_0(I-p~ y_ly_l^T)$ in one particular scenario based on supersymmetry. This form
is shown to lead to correct description of neutrino masses and mixing angles. The thermal leptogenesis after inclusion of flavour effects can account  for the observed baryon asymmetry of the universe within the present scenario. Rates for lepton flavour violating processes can occur at observable levels in the supersymmetric version of the scenario.
\end{abstract} 

\maketitle
\newpage 
\section{Introduction}
Our present knowledge \cite{rev} on neutrino parameters is based on information obtained from (a)  positive results of neutrino oscillation experiments (b) negative results of the neutrinoless double beta decay searches and (c) neutrino mass bounds from the 
cosmological observations.  Combinations of these allow two qualitatively different patterns for neutrino masses: One in which the neutrino masses follow some hierarchy, normal or inverted while in the other all three neutrino masses are (nearly) degenerate.
 The stringent constraints on the degenerate mass $m_0$ comes from cosmology. Depending on which data set one uses and method of analysis, 3$m_0$ can vary between 0.9-1.7 eV or 2-3 eV \cite{pastor}, the latter limit  is based solely on the information from the  cosmological microwave background studies.  All the neutrinos having a quasi degenerate mass in the range 0.3-1 eV is thus an allowed possibility.
It is non-trivial to accommodate this possibility  within the conventional pictures of neutrino mass generation. Indeed,  unified treatment of all fermion masses tend to generate hierarchical masses for neutrinos as well.  For example, the light neutrino masses are related to the Dirac neutrino mass matrix $m_D$ in  
 type I seesaw model \cite{rev}  and generically follow the hierarchical patterns.  The purpose of this letter is to identify symmetries and scenarios based on them which lead to quasi degenerate neutrinos  in  type I seesaw model which is by far the most popular mechanism for neutrino mass generation. Unlike all the  
previous models \cite{degm1, degm2} of degenerate neutrino masses,  
the existence of degeneracy in the present approach is insensitive to the detailed structure of $m_D$ and $M_R$ both of which can hierarchical and it is not linked to a type II contribution as is the case with some of the models \cite{degm1}.

Our basic formalism derives ideas from the minimal flavour violation (MFV) hypothesis \cite{mfv}  which uses  symmetries of the standard model Lagrangian to construct effective theories of flavour violations in frameworks going beyond it.  While the structure of flavour violations is not our primary concern we use similar ideas to constrain   structure of neutrino masses.
If quarks are massless then the SM Lagrangian is invariant under the flavour group ${\cal G}_q\equiv U(3)^3$ corresponding to independent unitary rotations on three flavours of quark doublets $q_L$, and singlets $u_R$ and $d_R$. The Yukawa couplings violate this symmetry preserving the baryon number and hypercharge symmetries. The basic assumption of MFV hypothesis is that these Yukawa couplings are the sole source of flavour violations and they determine the structure of flavour violations in theories which go beyond SM. 



MFV principle has also been used in the lepton sector as well in  several works \cite{grin1,davidson,hambye}. Its  implementation depends crucially on the source of neutrino masses and how lepton number gets violated.
Straightforward possibility \cite{grin1} assumes that  the charged lepton Yukawa couplings and the neutrino mass matrix appearing as coefficient of dimension 5 lepton number violating operator are the  irreducible sources of flavour violations and  an effective theory of flavour is constructed using them. The basic Yukawa couplings and the flavour symmetry would be different in more fundamental theory of neutrino masses. Consider for example the seesaw model with three right handed (RH) neutrinos $\nu_R$. The Yukawa couplings are given by \be \label{leptonic yukawa}
-{\cal L}_y=\bar{l}_L y_le_R\phi+\bar{l}_Ly_D\nu_R\tilde{\phi}+{\rm H.C.}~. \ee
$\phi$ is the standard Higgs doublet and $\tilde{\phi}=i\sigma_2\phi^*$. $y_l$ and $y_D$ are the Yukawa coupling matrices.
In the absence of these couplings, the Lagrangian is invariant under the symmetry group ${\cal G}_l\equiv U(3)_l\times U(3)_e\times U(3)_\nu$, where each $U(3)_f~(f=l,e,\nu)$,  corresponds to independent rotations on $l_L,e_R,\nu_R$.  Breaking of this flavour symmetry is governed by the Yukawa couplings as well as the  explicit Majorana mass term for the RH neutrinos
\be \label{mr}
\frac{1}{2}\nu_R^TC^{-1}M_R\nu_R~+~{\rm H. C.} \ee
The MFV hypothesis can be  implemented   by  assigning the Yukawa couplings $y_l$ and $y_D$ appropriate transformation property under ${\cal G}_l$  in such a way that eq.(\ref{leptonic yukawa}) becomes ${\cal G}_l$ invariant. 
The RH mass term is  not invariant under ${\cal G}_l$. It is made invariant in \cite{grin1} 
by assuming a smaller flavour symmetry group   $U(3)_l\times U(3)_e\times O(3)_\nu$ and assuming that $M_R$ is proportional to identity.  
We wish to consider here an alternative possibility which assumes  that $M_R$ also arises from presumably small $y_l,y_D$ in an effective theory from physics at high scale with  broken lepton number.  
This can be realized  if it is assumed that the relevant effective symmetry is a sub-group of ${\cal G}_l$. 
Specifically, we assume that \\

\noindent (1) The effective flavour symmetry group ${\cal G}_F\equiv O(3)_l\times O(3)_e\times O(3)_\nu\times U(1)_R$ of the Yukawa 
Lagrangian is a subgroup of the full flavour symmetry ${\cal G}_l$ of the SM Lagrangian without the  Yukawa couplings. The last $U(1)_R$ corresponds to the lepton number  transformation on the RH neutrinos.  
$y_l$ and $y_D$ are assumed to transform under ${\cal G}_F$ to make eq.(\ref{leptonic yukawa}) invariant under ${\cal G}_F$. Specifically,
\beqa \label{transformation}
l_L&\rightarrow& O_l \l_L~~,~~ e_R\rightarrow O_e e_R~~,~~ \nu_R\rightarrow O_\nu\nu_R
 ~~,\nonumber \\
y_l&\rightarrow& O_ly_lO_e^T~~~~~~,~~~~~~y_D\rightarrow O_ly_DO_\nu^T
~. \eeqa
$O_{l,e,\nu}$ are three orthogonal matrices.  In addition, $y_D$ and $\nu_R$ are assigned opposite charges under $U(1)_R$.
 We shall comment subsequently on other choices of ${\cal G}_F$. \\

\noindent  (2)  $y_l$ and $y_D$ are the only irreducible couplings which not only determine flavour violations but also the structure of 
$M_R$ and hence of  the neutrino mass matrix. $M_R$  is determined using $y_{l,D}$ and the above
transformation properties. 
\section{Effective $M_R$}
For orientation, we consider an explicit scheme to realize above assumptions. We introduce two complex fields $\eta_l$ and $\eta_D$.
They are singlets with respect to SM but transform under ${\cal G}_F$ respectively as 
$(3,3,1)_0$ and $(3,1,3)_{-1}$ where the suffix corresponds to $U(1)_R$ values. 
$\eta_l$ and $\eta_D$ are $3\times 3$ matrices in flavour space. The flavour symmetry forbids renormalizable Yukawa couplings 
but allows the following non-renormalizable operators as in  Froggatt Nielsen proposal \cite{fg}: 
\beqa \label{nr}
-{\cal L}_Y&=&
\frac{1}{2\Lambda}\nu_R^T C^{-1} \eta_D^T\left(c_0+\frac{c_1}{\Lambda^2}\eta_l\eta_l^T+\frac{d_1}{\Lambda^2}\eta_l^*\eta_l^\dagger+\frac{d_2}{\Lambda^2}(\eta_D\eta_D^\dagger+\eta_D^*\eta_D^T)+\frac{d_3}{\Lambda^2}(\eta_l \eta_l^\dagger+\eta_l^*\eta_l^T)+....\right)\eta_D\nu_R  ~,\nonumber \\
&+& 
\frac{1}{\Lambda}\left(\bar{l}_L\eta_l e_R\phi+\bar{l}_L\eta_D\nu_R\tilde{\phi}\right)+{\rm H. C.}~.  \eeqa
Here $c_0,c_1,d_1,d_2,d_3$ are coefficients of ${\cal O}$(1). 
Several comments are in order in connection with the above equation.
\begin{itemize}
\item Bare mass term for the RH neutrinos is not allowed by the $U(1)_R$ symmetry.
\item The RH neutrino mass term (first line of eq.(\ref{nr})) has to be symmetric in flavour space and its structure is completely determined by  ${\cal G}_F$ and the transformation rule given in eq.(\ref{transformation}). In particular, terms proportional to
$d_{1,2,3}$ would be absent from the superpotential of the supersymmetric generalization of the model.
They may arise at higher order from the D-terms.
\item  The total lepton number is explicitly broken in eq.(\ref{nr}) while the 
RH lepton number gets spontaneously broken by the vacuum expectation value of $\eta_D$. 
Lepton number conservation is restored  in the limit $\Lambda\rightarrow \infty $. $\Lambda$  therefore sets the scale of lepton number violation.
Flavour violations are determined by the scale $\equiv \Lambda_{FV}$ set by the vevs of  $\eta_{l,D}$ and 
$\Lambda_{FV}\lesssim\Lambda$.  $\vev{\eta_{l,D}}$ in fact play dual role here. On one hand, they determine the structure of the Yukawa couplings:
\be \label {ylyd} y_l=\frac{\vev{\eta_l}}{\Lambda}~,~y_D= \frac{\vev{\eta_D}}{\Lambda}~.\ee 
On the other, $\vev{\eta_{D}}$  also determines the RH handed neutrino masses:
\be \label{rhmasses} M_R\approx c_0\frac{\vev{\eta_D}^T\vev{\eta_D}}{\Lambda}~.\ee
This shows that the RH neutrino masses are suppressed compared to $\Lambda$ indicating the seesaw origin for these masses as well.
\end{itemize}
Neutrino masses follow from eqs.(\ref{nr},\ref{ylyd}):
\beqa\label{mdmr}
m_D&=&v y_D~,\nonumber \\
M_R&=&\Lambda y_D^T\left(c_0+c_1y_ly_l^T+d_1y_l^*y_l^\dagger +d_2(y_D y_D^\dagger+y_D^*y_D^T)+d_3(y_l y_l^\dagger+y_l^*y_l^T)\right)y_D
~,\eeqa
where $v\sim 174 $ GeV denotes the Higgs vacuum value. The light neutrino mass term is then given by:
$$\frac{1}{2}\nu_L^TC^{-1}{\cal M}_\nu^*\nu_L~+{\rm H.C.}~,$$
with 
\beqa \label{deg1}
{\cal M}_\nu&\equiv& m_DM_R^{-1}m_D^T ~, \nonumber \\
&\approx& m_0\left(1-\frac{c_1}{c_0}y_ly_l^T-\frac{d_1}{c_0}y_l^*y_l^\dagger-\frac{d_2}{c_0}(y_D y_D^\dagger+y_D^*y_D^T)
-\frac{d_3}{c_0}(y_l y_l^\dagger+y_l^*y_l^T)+....\right) ~.\eeqa
Seesaw mechanism together with ${\cal G}_F$ invariance  has resulted in an effective neutrino mass matrix with three
almost degenerate neutrinos having a common mass 
$$m_0\equiv\frac{v^2}{c_0\Lambda}~. $$

This is contrary to the standard expectations in the type-I seesaw model where hierarchical $m_D$ leads to hierarchical neutrino masses. 
The lepton number violation scale is restricted to be  $\Lambda \gtrsim10^{14}$ GeV for $c_0\sim 1$ and $m_0\lesssim 0.3 $eV. 
This scale could even be higher if $c_0$ is suppressed. Note that $m_0$ is independent of the scale of the flavour symmetry breaking
and the RH neutrino masses. This happens because of the seesaw origin of the RH neutrino masses, see eq.(\ref{rhmasses}). The role
of the RH neutrinos is to give a quasi degenerate spectrum through this double seesaw mechanism.  Moreover, the
${\cal G}_F$ invariance also results in a very specific structure of  departures
from degeneracy.  

The present scheme differs from all other previous models \cite{degm1,degm2} of the degenerate neutrinos in an important way. These models need to have some restrictions on the structure of the Dirac mass matrix and/or require
degenerate spectrum for the RH neutrino masses. In contrast, $m_D$ here can be arbitrary (as long as the Yukawa couplings $y_D<1$)
and it determines the structure of $M_R$. Both $m_D$ and $M_R$ can be simultaneously hierarchical yet result into (almost) degenerate spectrum after the seesaw mechanism once ${\cal G}_F$ invariance is imposed. 

The type I seesaw model and ${\cal G}_F$ invariance together led to quasi degeneracy. The quasi degeneracy  follows on a more general ground 
from the ${\cal G}_F$ invariance alone.  Consider SM model without RH neutrinos. Dirac Yukawa couplings $y_D$ are absent and appropriate symmetry would be $O(3)_l\times O(3)_e$ in this case. Neutrino masses can be understood as arising from an
 effective dimension five operator. Requiring invariance of this operator  under $O(3)_l\times O(3)_e$, the transformation rules eq.(\ref{transformation})  imply the same neutrino mass matrix as
in eq.(\ref{deg1}) but without the $y_D$ terms.

We close this section  with a comment on a possible origin of eq.(\ref{nr}) which was written  down in an effective theory approach using the MFV.  Let us add three sterile
neutrinos $s_R$ transforming under ${\cal G}_F$ as $(3,1,1)_0$. This allows the following renormalizable ${\cal G}_F$ invariant interactions
$$\sim \bar{s}_R^c\eta_D\nu_R+\frac{\Lambda}{2 } s_R^TC^{-1}s_R+{\rm H. C. } ~.$$
The bare mass term for $s_R$ sets the scale of lepton number violation.  Integration of $s_R$ after the seesaw mechanism generates the $\eta_D$-dependent terms 
given in the right handed mass matrix,  eq.(\ref{mdmr}).  \\ 

 \section{Neutrino masses and mixing}
From now onwards we specialize to supersymmetric model and consider the following simpler version of eq.(\ref{deg1}):
 \be \label{deg2}
{\cal M}_\nu\approx m_0(1-p \;y_l\;y_l^T)~,\ee
with $p\equiv \frac{c_1}{c_0}$ and now $m_0=\frac{v^2\sin^2\beta}{c_0\Lambda}$.
In this case, it is the charged lepton Yukawa couplings rather than $y_D$ which determine both mixing among neutrinos and their 
(mass)$^2$ differences.  This remarkably   simple structure  is capable of explaining all the features of the neutrino spectrum.  Necessary condition for this to happen 
is that either  $p$ and/or  $y_l$ are complex indicating the presence of CP violation in general. $p$ can be chosen real without loss of generality. $y_l$ also has fairly general structure in the absence of further assumptions in spite of some freedom offered by the ${\cal G}_F$ in the choice of flavour basis.  The structure of $y_{l,D}$ in the outlined model is determined by the vacuum structure of $\eta_{l,D}$ which together represent four independent real $3\times 3$ matrices.  ${\cal G}_F$ invariance can be used to make one of them (say $Re\vev{\eta_D}$ ) diagonal and the remaining freedom can be used to make three of the elements in $\vev{\eta_l}$ real or purely imaginary. This still allows fairly general
forms for $y_{l,D}$. Each choice of these couplings correspond to a specific direction in ${\cal G}_F$ space and implies a definite form for 
neutrino masses. 
Below we give specific but fairly general forms for $y_l$ leading to a successful description of the neutrino spectrum.

A general  $y_l$ can be written as
\be \label{yl}
y_l=V_{lL}d_lV_{lR}^\dagger ~.\ee
Here, $V_{lL,lR}$ are $3\times 3$ unitary matrices and $d_l={\rm diag.}(y_e,y_\mu,y_\tau)$ is the known diagonal matrix of the charged lepton Yukawa couplings, {\it e.g.} $y_\tau=\frac{m_\tau}{v\cos\beta}$ in the MSSM. 
Neutrino mixing is determined in the flavour basis defined as
\be \label{mnuf}
{\cal M}_{\nu f}\equiv V_{lL}^\dagger {\cal M}_\nu V_{lL}^*\approx m_0(U_{lL}-p\; d_l\;U_{lR}\;d_l )~,\ee
where $U_{lL,lR}\equiv V_{lL,lR}^\dagger V_{lL,lR}^*$ are symmetric  unitary matrices. In particular, if $y_l$ are real then $V_{lL,lR}$
are orthogonal and  $U_{lL}=U_{lR}=1$.  ${\cal M}_{\nu f}$ becomes diagonal in this case and there is no mixing among
neutrinos although they are now non-degenerate.  Thus we need to assume $V_{lR}$ and/or $V_{lL}$ to be  complex. 
The first term corresponds to the most general mass matrix for the degenerate
neutrinos studied in \cite{branco} . $U_{lL,lR}$ are both symmetric and unitary and can be parametrized \cite{branco} as
\be \label {param}
U_{lL,lR}=P_{L,R}R_{23}^T(\phi_{L,R})U_{12}(\theta_{L,R},\alpha_{L,R})R_{23}(\phi_{L,R})P_{L,R} ~,\ee
where 
\be \ba{cc}
R_{23}(\phi)=\left( \ba{ccc}
1&0&0\\
0&\cos\phi&\sin\phi\\
0&\sin\phi&-\cos\phi \\  \ea \right)
&;U_{12}(\theta,\alpha)=\left( \ba{ccc}
\cos\theta&\sin\theta&0\\
\sin\theta&-\cos\theta&0\\
0&0&e^{i\alpha} \\  \ea \right)
\ea ~\ee
and $P_{L,R}$ are diagonal phase matrices. Phases in one of these can be removed by redefining the phases of the charge leptons, see eq.(\ref{mnuf}) and we choose to make $P_L=I$.

Notice that $U_L$ becomes invariant under $\nu_\mu\leftrightarrow \nu_\tau$ interchange if $\phi_L=\frac{\pi}{4}$. 
This makes ${\cal M}_{\nu f}$ \mt symmetric to leading order and leads to  prediction \cite{rev}  of the  maximal atmospheric mixing  angle $\theta_{23}$ and vanishing $\theta_{13}$. The second term in eq.(\ref{mnuf}) generically violates \mt symmetry by a small amount  since $y_\mu\not= y_\tau\ll1$. But even in this case, there is a unique choice for $U_{lR}$ which allows \mt symmetric perturbations as well. This is given by
\begin{equation} \label{ulr}
U_{lR}=\left( \ba{ccc}
1&0&0\\
0&0&e^{i\beta_3} \\
0&e^{i\beta_3}&0\\
 \end{array} \right) ~. 
\end{equation}

For this special case, the departure from degeneracy are characterized by a single parameter $\epsilon_1\equiv p\;y_\mu\;y_\tau$
and the solar scale $\Delta_\odot$, the atmospheric scale $\Delta_A$  and the solar angle $\theta_{12}$ get determined as  
$\epsilon_1$ as:
\beqa \label{predict}
\tan 2\theta_{12}&\approx&\frac{\tan\theta_L}{\cos\beta_3}\left(1-\frac{\epsilon_1}{2}\frac{\sin^2\beta_3\cos\beta_3}{\cos\theta_L}+{\cal O}(\epsilon_1^2)\right)~, \nonumber \\
\Delta_\odot \cos2 \theta_{12}&\approx&2 m_0^2 \epsilon_1\left( \cos\theta_L\cos\beta_3+{\cal O}(\epsilon_1)\right)~,\nonumber \\
\Delta_A&\approx&2 m_0^2  \epsilon_1 \left(\cos(\alpha_L-\beta_3)+\frac{\cos\beta_3\cos\theta_L \sin^2\theta_{12}}{\cos 2 \theta_{12}}+{\cal O}(\epsilon_1)\right)~.\eeqa
Note that the solar mixing angle arises at zeroth order and perturbation makes only a small change. The $\Delta_\odot,\Delta_A$
are generated at the first order in $\epsilon_1$.   Above equations serve to determine $\beta_3,m_0^2 \epsilon_1$ in terms of $\theta_L$ and the  known quantities. $\theta_L$ is required to be small and $\beta_3$ close to $\frac{\pi}{2}$ for obtaining the correct $\theta_{12}$ and $\frac{\Delta_\odot}{\Delta_A}$.  Similarly, $\alpha_L$ is also
required to be non-zero. Both these phases violate CP and appear as phases in the Majorana mass for light neutrinos. Underlying CP violation is however not manifested
through the Dirac phase because of the \mt symmetry which makes $\theta_{13}$ zero.  We discuss below a generalization which does not impose \mt symmetry and leads to 
observable CP violating phase.

In the most general  situation and neglecting $y_e$ , perturbations to degeneracy are governed by two parameters $\epsilon_1$ defined above and $\epsilon_2\equiv p y_\tau^2$. It follows from eq.(\ref{mnuf}) that $\epsilon_2$ would dominate except when $(U_{lR})_{33}\ll \frac{y_\mu}{y_\tau} (U_{lR})_{22}$. Specific choice $(U_{lR})_{33}\approx 0$ (corresponding to $\cos^2\phi_R e^{i \alpha_R}\approx \sin^2\phi_R \cos\theta_R$) in eq.(\ref{param}) is quite interesting. In this case, the lower $2\times 2$ block of ${\cal M}_{\nu f} {\cal M}_{\nu f}^\dagger $ becomes \mt symmetric  (up to terms of order $\frac{y_\mu}{y_\tau}$) for arbitrary values of other angles and phases in $U_{lL,lR}$. Thus this choice naturally leads to
a large atmospheric mixing angle. Moreover, perturbations are essentially controlled in this case by the single parameter $\epsilon_1$.

\begin{figure}[t]
 \centering
 \includegraphics[width=11cm,height=7cm,bb=0 0 420 271]{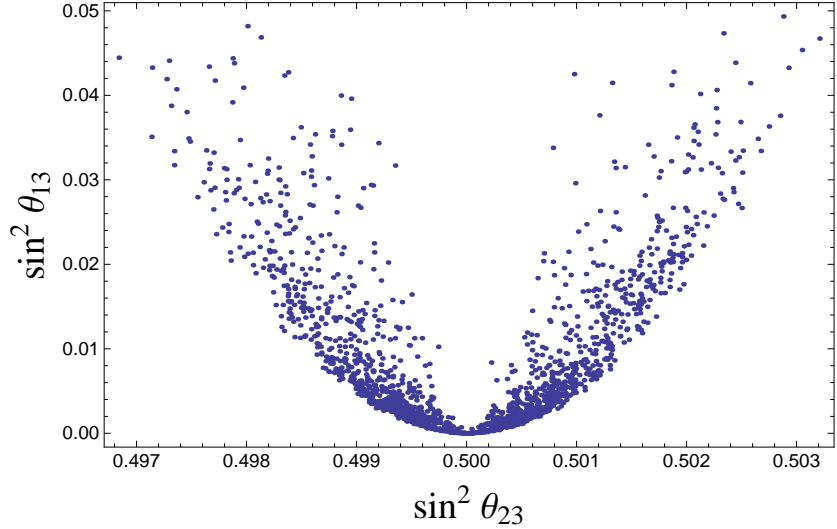}
\caption{Allowed ranges of $\sin^2\theta_{23}$ versus $\sin^2\theta_{13}$ in model implied by eq.(\ref{mnuf}) and the assumption $(U_{lR})_{33}=0$, see text for details}
\end{figure}

\begin{figure}
 \centering
 \includegraphics[width=11cm,height=7cm,bb=0 0 427 270]{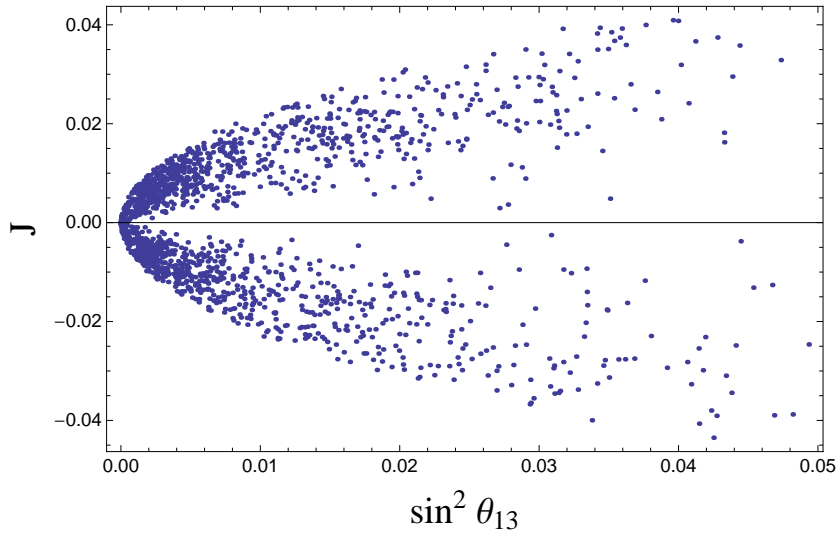}
\caption{Allowed ranges of $\sin^2\theta_{13}$ versus the Jarlskog invariant $J$  in model implied by eq.(\ref{mnuf}) and the assumption $(U_{lR})_{33}=0$, see text for details}
\end{figure}

We have numerically studied implications of the specific choice $(U_{lR})_{33}=0$ using the general parametrization in eq.(\ref{param}). We randomly vary the independent parameters 
$\theta_L,\phi_L,\beta_{3},\alpha_L$ in the full range. $\beta_{3}$ here represents the phase of $(U_{lR})_{23}$ in eq.(\ref{param}).
$\epsilon_1$ is redefined  as   $\epsilon_1\equiv p\;y_\mu\;y_\tau |(U_{lR})_{23}|$.  Since entire mixing and mass differences are determined by the perturbations in eq.(\ref{mnuf}), $m_0^2 \epsilon_1$ acts as a normalization constraint but we have
imposed the condition that $\epsilon_1<0.5$ and $m_0<0.3$ eV.
 We require that each choice of random inputs correctly reproduce the $\Delta_\odot,\Delta_A,\sin^2 2\theta_{23},\tan^2\theta_{12}$
within $1\sigma$.  Two specific outcomes of this random analysis are displayed in Fig.(1) and Fig.(2). Fig.(1) displays the variation of $\sin^2\theta_{13}$ with $\sin^2 \theta_{23}$. Interestingly, the atmospheric mixing stays close to maximal for all of the random allowed choices mentioned above but the $\theta_{13}$ can span the entire $3\sigma$ range. CP violation is an essential ingredient in this analysis since its absence implies no mixing as argued above. Fig.(2) displays the allowed values of the Jarlskog invariant $J$ versus
$\sin^2\theta_{13}$. There is a clear correlation between $J$ and the $\sin^2\theta_{13}$ which can be tested.
The above results hold for strictly zero $(U_{lR})_{33}$. We find that one could obtain sizable deviation from the maximality by allowing 
$(U_{lR})_{33}\sim \frac{y_\mu}{y_\tau} (U_{lR})_{22}$. Even in this case, large values of  $\theta_{13}$ is possible and prediction of the Jarlskog invariant shown in Fig.(2) does not change appreciably.

The above discussion was based on the specific choice ${\cal G}_F=O(3)_l\times O(3)_e\times O(3)\nu\times U(1)_R$. One could consider
a smaller symmetry by replacing $O(3)_l\times O(3)_e$ by its vectorial sub-group under which both $l,e_R$ transform as triplets.
Quasi degeneracy is still maintained but now eq.(\ref{deg2}) gets replaced by
 \be \label{deg3}
{\cal M}_\nu\approx m_0(1-p( y_l+y_l^T)+....)~,\ee
Now the departure from degeneracy occurs at first order in the Yukawa couplings. Just as in the previous case this case  too is capable of reproducing the neutrino spectrum.

A larger choice for ${\cal G}_F$ can also lead to degenerate spectrum with additional assumptions. Consider the group
${\cal G}_F=U(3)_l\times U(3)_e\times O(3)_\nu\times U(1)_R$. This is the group  chosen in ref \cite{grin1} except for the additional $U(1)_R$. Using the same arguments as in the above cases, leading order
expression of $M_R$ is given by
$$M_R=c_0\Lambda (y_D^\dagger y_D+y_D^Ty_D^*)+.....$$
This $M_R$ also leads to degenerate neutrinos provided $y_D^\dagger y_D$ is real.\\
  
\section{ Leptogenesis}
The baryon asymmetry $Y_B$ in the universe can be generated through leptogenesis \cite{nir} in the present approach after due considerations of flavour effects \cite{flv1,flv2,nir}. The latter become important in MSSM for temperature below $(1+\tan^2\beta)  10^{12}$ GeV when the tau interactions
start equilibrating and tend to wash out asymmetry in the tau flavour.  In a general seesaw framework, this asymmetry depends on two distinct set of parameters: The Dirac Yukawa couplings and the right handed neutrino masses. Here, the single relevant quantity is
the Dirac Yukawa coupling matrix $\tilde{y}_D$ in basis with diagonal charged leptons and the RH neutrinos.
Let $V_R$ be a unitary matrix which diagonalizes $M_R$ in eq.(\ref{mdmr}).
$$V_R^T M_R V_R=D_R~,$$
where $D_R$ is diagonal matrix of the RH neutrino masses. $\tilde{y}_D$ is then given by
$$\tilde{y}_D=V_{lL}^\dagger  y_DV_R ~.$$

The lepton asymmetry generated in flavour $\alpha=~(e,\mu,\tau)$ by the out of equilibrium decay of the lightest right handed neutrino is given in the MSSM by \cite{nir}
\be \label{epsalpha}
\epsilon_{\alpha\alpha}\approx -\frac{3M_1}{8 \pi M_k} \frac{ Im[(\tilde{y}_D^\dagger \tilde{y}_D)_{1k}(\tilde{y}_D^\dagger)_{1\alpha} (\tilde{y}_D)_{\alpha k}]}{(\tilde{y}_D^\dagger \tilde{y}_D)_{11}} ~,\ee
where $M_k\;,{k=1,2,3}$ represent the RH neutrino masses and $M_1$ corresponds to the lightest one.
The $Y_B$ generated through $\epsilon_{\alpha\alpha}$ also depend on the wash out parameters
\be \label{washout1}
\tilde{m}_\alpha=\frac{v^2\sin^2\beta(\tilde{y}_D^\dagger)_{1\alpha}(\tilde{y}_D)_{\alpha 1}}{M_1} ~. \ee

 Eq.(\ref{mdmr}) can be written in terms of $\tilde{y}_D$ as
\be \label{dr}
D_R=c_0 \Lambda \tilde{y}_D^T V_{lL}^T(1+ p~ y_ly_l^T)V_{lL}\tilde{y}_D~, \ee
We can invert above equation to obtain a parametrization for $\tilde{y}_D$\cite{ibarra}:
\be \label{ytilde}
\tilde{y}_D\approx V_{lL}^\dagger(1-\frac{1}{2} p~ y_ly_l^T)R \left(\frac{D_R}{c_0\Lambda}\right)^{1/2}\ee
to leading order in $p~y_ly_l^T$. Here $R$ is a complex orthogonal matrix.  

The expressions for $\epsilon_{\alpha\alpha}$ get simplified in the \mt symmetric limit, $ \phi_L=\frac{\pi}{4}$,  in eq.(\ref{param})
In this limit, $V_{lL}$ which reproduces $U_{lL}$ in  eq.(\ref{param}) is given by \cite{branco}
\be \label{vl}
V_{lL}\approx \left(\ba{ccc}
1&0&0\\
0&-\frac{i}{\sqrt{2}}&-\frac{i}{\sqrt{2}}\\
0&\frac{1}{\sqrt{2}}e^{-i\alpha_L/2}&-\frac{1}{\sqrt{2}}e^{-i\alpha_L/2}\\
\ea \right) ~.\ee

It is now straightforward to work out the lepton asymmetries and wash out parameters:
\beqa \label{epsilon2}
\epsilon_e&\approx& -\frac{3M_1m_0}{8 \pi v^2\sin^2\beta}~\frac{Im [R_{11}^{*2}]}{(R^\dagger R)_{11}}~,\nonumber \\
\epsilon_\mu&\approx& -\frac{3M_1m_0}{16 \pi v^2\sin^2\beta}~\frac{Im [(R_{12}^\dagger-i R^\dagger_{13}e^{-i\alpha_L/2})
(R_{12}^\dagger+i R^\dagger_{13}e^{i\alpha_L/2})]}{(R^\dagger R)_{11}}~,\nonumber \\
\epsilon_\tau&\approx& -\frac{3M_1m_0}{16 \pi v^2\sin^2\beta}~\frac{Im [(R_{12}^\dagger+i R^\dagger_{13}e^{-i\alpha_L/2})
(R_{12}^\dagger-i R^\dagger_{13}e^{i\alpha_L/2})]}{(R^\dagger R)_{11}}~.\eeqa
\be
\tilde{m}_e \approx m_0 |R_{11}|^2~;~ \tilde{m}_\mu\approx \frac{m_0}{2} |R_{21}-iR_{31} e^{i \alpha_L/2}|^2~;~ 
\tilde{m}_\tau\approx \frac{m_0}{2} |R_{21}+i R_{31} e^{i \alpha_L/2}|^2 ~.\ee
In writing above equations, we have  retained only  zeroth order terms in $p$. 
Note that the scale of the individual lepton asymmetries is set by the degenerate mass $m_0$ and not the atmospheric or solar scale as in models with hierarchical neutrinos. Also it is easy to check from the leading order expression given above that the sum $\epsilon_l\equiv\sum_{\alpha}\epsilon_{\alpha\alpha}$ vanishes. By including the non-leading terms and using eq.(\ref{predict}) , we find
\be  \label{el}
\epsilon_l\approx  \frac{3M_1^2}{8\pi v^4} \Delta_\odot \cos 2 \theta_{12}\frac{Im [R_{11}^{*2}]-2 \sin\frac{\alpha_L}{2} Im [R_{21}^*R_{31}^*]}{(R^\dagger R)_{11}} ~.\ee 
This is much smaller than the individual asymmetries. The latter can be quite large $\sim {\cal O}( 10^{-5})$ for $M_1\sim 10^{10}$ GeV.
We use the approximate approximation for the $Y_B$ given in \cite{flv2}:
\be Y_B\approx -\frac{10}{31 g_*}\left[ \epsilon_e \eta\left( \frac {93}{110} \tilde{m}_e\right)+\epsilon_\mu \eta\left( \frac {19}{30} \tilde{m}_\mu\right)+\epsilon_\tau \eta\left( \frac {19}{30} \tilde{m}_\tau\right)\right] ~\ee
valid for the temperature range $(1+\tan^2\beta) 10^5~{\rm GeV}\leq T\leq (1+\tan^2\beta) 10^{9}~{\rm GeV}$.  The washout function $\eta(x) $ is given by
$$ \eta(x)\approx \left[ \left(\frac{x}{8.25\times 10^{-3}~{\rm eV}}\right)^{-1} +\left(\frac{0.2 \times10^{-3}~{\rm eV}}{x}\right)^{-1.16}\right]^{-1}~$$
and $g_*=228.75$.
Note that the matrix $R$ is independent of the light neutrino mass parameters and there exists large ranges of three complex angles parametrized $R$ for which $Y_B$ can be significant. We give one set for illustrative purpose. Let us define
$R_{11}=\cos z_{13}\cos z_{12}~,~ R_{21}=\cos z_{13}\sin z_{12}$ and $R_{31}=\sin z_{13}$ where $z_{ij}$ are complex angles.
Then the choice $z_{13}=0 , z_{12}=0.30+ 0.13 I , m_0=0.3 ~{\rm eV} , M_1=7.9\times 10^{10}~ {\rm GeV}, \tan\beta=10$ leads to
$Y_B\approx 8.6\times 10^{-11}$. Individual lepton asymmetries are quite large for this choice $\epsilon_\mu=\epsilon_\tau=-\frac{1}{2} \epsilon_e\approx 7.4 \times 10^{-6}$ but there sum vanishes emphasizing the role 
played by flavour effects.\\ 

\section{Lepton Flavour Violation}
As in other MFV approaches, the structure of the leptonic flavour violations is coded in $y_l,y_D$.
The ratio of the scale of lepton number violation , $\Lambda$ to lepton flavour violation $<\eta_D>$ is not  
very large $\sim (1- 100)$. Consequently, if these are the only scales in the theory, then the the rates of lepton flavour violating (LFV) processes like $l_i \to l_j + \gamma$ are highly suppressed (see for example, the discussion in Ref. \cite{grin1}). 

In the supersymmetric version of the theory however, there is a new scale at  low energy in terms of the slepton and sneutrino masses at the weak scale. These soft masses
continue to carry the memory of high scale flavour violation due to the presence of the seesaw mechanism, leading to large flavour violating effects at the
weak scale\cite{standardlfv}. In our present scheme, we assume that the soft masses are universal below the high scale $\Lambda$. This sets the following 
hierarchy of scales $$ \Lambda ~\gtrsim ~\Lambda_{\tilde{m}_0} ~\gtrsim ~<\eta_D> ~\gtrsim M_R,$$ where $\Lambda_{\tilde{m}_0}$ determines the scale
where soft masses are universal. At the weak scale, the sleptons receive corrections proportional to $y_D, y_l$ due to renormalization group (RG) effects, which 
are roughly given as
\begin{eqnarray}
m_{\tilde{L}}^2  &\approx& \tilde{m}_0^2 \left( k_0  I - y_D y_D^\dagger l_1^0 - y_l y_l^\dagger l_2^0  \right) \nonumber \\ 
m_{\tilde{e^c}}^2 &\approx & \tilde{m}_0^2 \left( k'_0 I - y_l^\dagger y_l  l_3^0 \right) ,
\end{eqnarray}  
where $l_1^0,l_2^0,l_3^0, k_0,k'_0$ are coefficients generated by RG running with a typical size of the order $1/(16 \pi^2 ) \ln \Lambda^2/\tilde{m}^2_0$.
In the basis where charged leptons and the RH neutrinos are diagonal, the flavour off-diagonal entries in the slepton mass matrices determine the amplitudes of the flavour
violating processes. In this basis, these off-diagonal entries are proportional to $\tilde{y}_D \tilde{y}_D^\dagger$ with $\tilde{y}_D$ given by 
eq.(\ref{ytilde}).  At the leading order  this takes the form :
\begin{equation}
 \tilde{y}_D \tilde{y}_D^\dagger \approx V_{lL}^\dagger R ~{ D_R  \over c_0 \Lambda } ~R^\dagger V_{lL}  + \mathcal{O}(p~y_ly_l^{T}) ,
 \end{equation}
where we have neglected $\mathcal{O}(p~ y_l y_l^{T})$ corrections. The strength of flavour violation is best judged by considering the ratio of
the flavour violating off-diagonal entries to the flavour diagonal terms. Here we have 
\begin{equation}
\label{deltas}
(\delta^{(l)}_{LL})_{ij}|_{i \neq j}  \approx {m_0 \over v^2} \left[ V_{lL}^\dagger R D_R \tilde{t}_0 R^{\dagger} V_{lL} \right]_{ij}
\end{equation}
 Here $i,j$ are generation indices and $\tilde{t}_0$ is a diagonal matrix containing  the logarithmic terms for each of the right handed neutrinos, 
 given as $1/(16 \pi^2) \ln <\eta_D>^2/M_i^2$. We have also exchanged $c_0 \Lambda$ for the light neutrino mass scale $m_0$.  Note that
 when $D_R$ is degenerate (or $y_D y_{D}^T = I$ ) and $R$ is real, there is no flavour violation in the theory. Assuming that the
 mass scale of the right handed neutrinos is roughly the same, we have the log factor to be $\ln <\eta_D>^2 /M^2~ \approx~ \ln \Lambda^2/<\eta_D>^2$.
Using eq.(\ref{deltas}) one can estimate the branching fraction of the LFV process $l_j \to l_i + \gamma$ as\cite{hisano} 
\begin{equation}
Br ( l_j \to l_i  + \gamma) \approx { \alpha^3 \over G_F^2 } { |(\delta_{LL}^{(l)})_{ij}|^2 \over m_{susy}^4 } \tan^2\beta 
\end{equation}
 Existing limits on $\mu \to e + \gamma$ from the MEGA experiment constraint $ |(\delta_{LL}^{(l)})_{ij}| \lesssim 10^{-4}$ for slepton masses $\sim~ m_{susy}~ \approx 400 $ GeV
 and tan$\beta~ \sim 10$ \cite{npb}.   Presently, the range between $10^{-6}- 10^{-4}$ in
  $(\delta^{(l)}_{LL})_{21}$ is being probed.  
 The explicit form of the relevant $(\delta_{LL}^{(l)})_{ij}$ in this case is given by 
 \begin{eqnarray}
 (\delta^{(l)}_{LL})_{21} & =& {m_0 \over \sqrt{2}  v^2}  \left[ (-i R_{11} R_{21}^\star + e^{-i \alpha_L/2}  R_{11} R_{31}^\star) M_1 (\tilde{t}_0)_{11} \right. \nonumber \\
 &- & \left. ( i R_{12} R_{22}^\star - e^{-i \alpha_L/2 } R_{12} R_{32}^\star ) M_2  (\tilde{t}_0)_{22 } -  
  ( i R_{13} R_{23}^\star - e^{-i \alpha_L/2 } R_{13} R_{33}^\star ) M_3  (\tilde{t}_0)_{33} \right] 
       \end{eqnarray} 
From the above we see that for $m_0 \sim 0.1 ~\text{eV}$ and $M_3 ~\sim10^{14} \text{GeV}$ ,  $(\delta^{(l)}_{LL})_{21}$ is close to $\mathcal{O}(1)$. This will require additional suppression from the elements of $R$ matrix,e.g. $R_{13}\sim 10^{-3}-10^{-4}$ can suppress the $(\delta^{(l)}_{LL})_{21}$. The leptogenesis can still work as exemplified above. Nevertheless both leptogenesis and LFV can together provide tight constraints on the model parameter space and it would be interesting to pursue this further.

\section{ Summary}
We have discussed here a novel approach to obtaining quasi-degenerate neutrinos in the context of type-I seesaw models. This is based on generalization of the 
minimal flavour violation  principle to the leptonic sector.  Earlier attempts in this direction regarded $M_R$ as an independent entity proportional to identity.
Instead if it is assumed that $M_R$ also arises from the Dirac Yukawa couplings in an effective theory then this can lead to quasi degenerate neutrino. As discussed here, this is possible by requiring invariance under ${\cal G}_F=O(3)_l\times O(3)_e\times  O(3)_\nu \times U(1)_R$. Larger choice for ${\cal G}_F$ is also shown to lead to degeneracy with additional assumption.  Consequences of the quasi degenerate structure implied by ${\cal G}_F$ were worked out. CP violation was found to be necessary in order to obtain
non-trivial mixing in this approach. Allowing this, one can obtain correct mixing patterns and observable CP violation. Thermal leptogenesis aided by flavour effects is shown to explain the observed baryon asymmetry. If CP violation is found to be small or nearly absent in future neutrino oscillation experiments then this approach will be strongly constrained if not ruled out entirely.\\

{\bf Acknowledgment:} ASJ would like to thank the Centre for High Energy Physics for hospitality where this work was initiated. SKV would like to acknowledge
support from DST Ramanujan Fellowship SR/S2/RJN-25/2008 from Govt. of India.

\end{document}